\documentclass[twocolumn,showpacs,aps,amssymb,floatfix,prd,amsmath,preprintnumbers]{revtex4}
\usepackage{epstopdf}
\usepackage{capt-of}
\usepackage{graphicx,latexsym}  
\usepackage{amssymb,amsmath}
\usepackage{amsthm}
\usepackage{url}
\usepackage{dcolumn}   
\usepackage{bm}
\DeclareMathOperator{\sech}{sech}
\begin{document}
\title{Cosmology from an exponential dependence on the trace of the energy-momentum tensor -  Numerical approach and cosmological tests}\

\author{
    G. Ribeiro$^1$, R. Sfair$^1$, P.H.R.S. Moraes$^{2}$, J.R.L. Santos$^3$, A. de Souza Dutra$^1$}
    \affiliation{$^1$UNESP - Universidade Estadual Paulista ``J\'ulio de Mesquita Filho" - Departamento de F\'isica e Qu\'imica, 12516-140, Guaratinguet\'a, S\~ao Paulo, Brazil}
    \affiliation{$^2$ITA - Instituto Tecnol\'ogico de Aeron\'autica - Departamento de F\'isica, 12228-900, S\~ao Jos\'e dos Campos, S\~ao Paulo, Brazil}
	\affiliation{$^3$Unidade Acad\^emica de F\'isica, Universidade Federal de Campina Grande, 58429-900 Campina Grande, PB, Brazil}

\begin{abstract}

In this paper, we present the cosmological scenario obtained from $f(R,T)$ gravity by using an exponential dependence on the trace of the energy-momentum tensor. With a numerical approach applied to the equations of motion, we show several precise fits and the respective cosmological consequences. As a matter of completeness, we also analyzed cosmological scenarios where this new version of $f(R,T)$ is coupled with a real scalar field. In order to find analytical cosmological parameters, we used a slow-roll approximation for the evolution of the scalar field. This approximation allowed us to derived the Hubble and the deceleration parameters whose time evolutions describe the actual phase of accelerated expansion, and corroborate with our numerical investigations. Therefore, the analytical parameters unveil the viability of this proposal for $f(R,T)$ in the presence of an inflaton field.  

\end{abstract}

\pacs{04.50.Kd; 98.80.Cq;  02.60.-x; 04.20.Jb.}


\keywords{$f(R,T)$ gravity; cosmology; numerical solutions; cosmological tests}

\maketitle


\section{Introduction}

	Einstein's theory of General Relativity (GR) has made a great impact on modern physics. The recent detection of gravitational waves \cite{abbott/2016,abbott/2017} is a strong proof that GR is a great tool to understand the behavior of gravitation in the Universe. 
   
  With modern cosmological observations, we concluded that our Universe is currently passing through a phase of accelerated expansion \cite{weinberg/2013,moresco/2016,schrabback/2010}. Within standard $\Lambda$CDM model of cosmology, the composition of the Universe is essentially $\sim5\% $ baryonic matter, $\sim25\% $ dark matter and $\sim70\%$ dark energy (DE), 
   with the DE fluid being capable of predicting the non-intuitive cosmic acceleration \cite{perlmutter/1999,riess/1998}. In fact, the $\Lambda$CDM model was introduced in order to explain this phenomenon \cite{peebles/2003}, for which $\Lambda$ is the cosmological constant inserted in the field equations of GR. However, the same cosmological constant when is evaluated from a particle physics approach expresses a value many orders of magnitude different \cite{weinberg/1989}, which leads us to search for some alternative path to match the current observations.
   
  Thereby, in recent works, theoretical physicists have chosen to modify GR to approach the cosmic acceleration \cite{alfaro/2012,alfaro/2013,wesson/1990,moraes/2016,magueijo/2003,moffat/2016}. 
  Not only the late acceleration phenomenon has been faced as a motivation for such new investigations but also the unification with the inflationary primordial Universe \cite{nojiri/2017}. 
   
  In the present paper, we will introduce a further alternative model to approach the cosmic acceleration in which the modification comes fundamentally at the gravitational action of GR. In our approach, the usual linear dependence on the Ricci scalar $R$ found in Einstein-Hilbert gravitational action is replaced by a general relation of $R$ and $T$, with the latter being the trace of the energy-momentum tensor \cite{harko/2011}. Such a generalization generates extra terms in the field equations of the model when compared to GR field equations and those extra terms may, in principle, attend to some cosmological observed features. There are other efficient generalizations of GR, suchlike $f(R)$ \cite{de_felice/2010}, $f(\mathcal{G})$ \cite{nojiri02/2005} and $f(\mathcal{L}_m)$ theories \cite{bertolami/2008}, for which $\mathcal{G}$ stands for the invariant topological scalar of Gauss-Bonnet and $\mathcal{L}_m$ is the matter-energy Lagrangian.
     
  The application of these alternative theories in astrophysical and cosmological phenomena has shown excellent matching results with current observations. For instance, by using a quadratic polynomial function of $R$ in the gravitational action, known as the Starobinsky model, it is possible to describe the early inflationary phase of the Universe \cite{paliathanasis/2017}. Also, studies on wormholes \cite{sharif-rani/2014} and stellar models \cite{bhatti/2017} were made under extended gravity. The application of $f(R,T)$ gravity turns wide open not only the use of the Friedmann-Lemaître-Robertson-Walker cosmological metric \cite{singh/2016,ms/2017} but also the Bianchi-type metric, which describes a homogeneous and anisotropic fluid permeating the Universe \cite{reddy/2012}. In such a theory, one can also find the analysis of the G\"{o}del metric, known as the simplest metric that allows closed time-like curves or time travels \cite{santos/2013}. 

  In this article, we intend to further contribute within $f(R,T)$ gravity framework. We will investigate the cosmological features of the $f(R,T)$ gravity by taking $f(R,T)=R+ \alpha e^{\beta T}$, with $\alpha$ and $\beta$ being constants. In Section II we present the $f(R,T)=R+\alpha e^{\,\beta\,T}$ gravity formalism. We analyze the cosmological features in a matter-dominated scenario using the flat Friedmann-Lemaître-Robertson-Walker metric. Some cosmological tests are applied to our model in Section III. Finally, in Section IV we present the discussion of our results and the final conclusions of this work. 
	
\section{Cosmological features of an exponential $f(R,T)$ model}

	As initially proposed in \cite{harko/2011}, the total action of the $f(R,T)$ gravity is expressed by
	
\begin{equation}\label{sgrav}
S=\frac{1}{16\pi}\int f(R,T)\sqrt{-g}d^4x +\int \mathcal{L}_m \sqrt{-g}d^4x.
\end{equation}
In \eqref{sgrav}, $g$ is the determinant of the metric tensor $g_{\mu\nu}$. We also consider the speed of light $c$ and the Newtonian gravitational constant $G$ as $c=G=1$ through this article for reasons of suitable simplifications.
	
	By taking $f(R,T)=R+\alpha e^{\,\beta T}$, the variation of \eqref{sgrav} with respect to the metric tensor is expressed as:
	
\begin{equation}\label{fe}
G_{\mu\nu}=8\pi T_{\mu\nu}+\alpha e^{\beta T}\left[\frac{1}{2}g_{\mu\nu}+\beta(T_{\mu\nu}+pg_{\mu\nu})\right]
\end{equation}
in which the recovery of GR is clearly observed for the case $\alpha=0$. 

	We shall consider a flat, homogeneous and isotropic Friedmann-Lemaître-Robertson-Walker metric in order to construct our cosmological model, so that
		
\begin{equation}\label{ds}
ds^2=dt^2-a^2(t)(dx^2+dy^2+dz^2),
\end{equation}
where $a(t)$ is the scale factor, which dictates how distances evolve during the Universe evolution.

	By substituting \eqref{ds} in \eqref{fe} yields
	
\begin{eqnarray} 
3\left(\frac{\dot{a}}{a}\right)^2=8\pi \rho +\alpha e^{\beta (\rho -3p)}\left[\beta(\rho -3p)+\frac{1}{2}\right], \label{00} \\
2\frac{\ddot{a}}{a}+\left(\frac{\dot{a}}{a}\right)^2=-8\pi p+\frac{1}{2}\alpha e^{\beta (\rho -3p)}. \label{11}
\end{eqnarray}
Besides, in the above equations we have also considered the energy-momentum tensor of a perfect fluid.

In order to solve Eqs.\eqref{00}-\eqref{11} we will invoke an equation of state like $p=\omega\rho$, with $\omega$ being the equation of state parameter. In fact, we will go further. We desire to check if the extra terms in such an exponential model of gravity are capable of predicting the cosmic acceleration with no need for invoking the existence of an exotic fluid, such as dark energy. To say that there is no dark energy (or phantom fluid \cite{nojiri01/2005}, quintessence \cite{amendola/2000,chiba/2000}) in the Universe today is to say that $\omega=0$ in $p=\omega\rho$, that is, non-relativistic pressureless matter dominates the Universe dynamics today. By taking $\omega=0$ in \eqref{00}-\eqref{11}, let us check if it is still possible to predict an accelerated regime for the Universe expansion.
\subsection{Cosmological solutions}	
	
\begin{figure}[h!]
\vspace{0.3cm}
\centering
\includegraphics[height=5cm,angle=00]{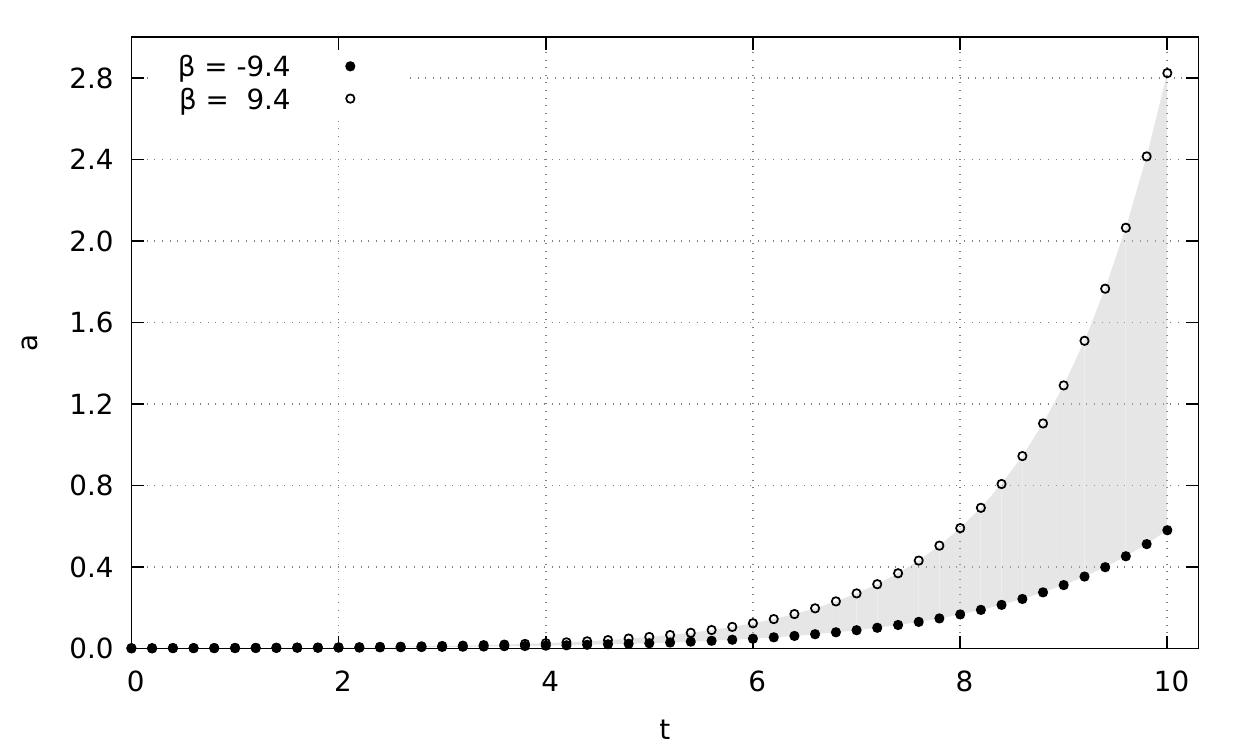}
\caption{Time evolution of the scale factor numerical solution, where the white dots were obtained for $\beta=9.4$, while the black dots were obtained for $\beta=-9.4$.}
\label{NumPlot_a}
\end{figure}

	  The condition $\omega =0$ yields to simplified versions of the equations \eqref{00} and \eqref{11} so that, with a straightforward algebra it is possible to reach a differential equation for $a(t)$. However, this previous differential equation turns to be highly non-linear and solving it analytically becomes infeasible. In this way, it is convenient to use numerical solving methods. By applying a data table for the numerical solution $a(t)$, where $t$ is running with intervals of $0.2$, into a graphic, it is possible to fit an appropriate arbitrary function that contains the referred values. In this way, it is possible to obtain the time evolution of the scale factor in accordance with Fig.\ref{NumPlot_a}.
	  
	The limited shaded area represents values in which $a(t)$ is mathematically valid, with $\beta\in[-9.4,+9.4]$. The constant $\alpha=1$ is fixed for all range, since for other values the solution may not occur. 
	
	Therefore our goal now turns out to be finding out an appropriate fitting function that expresses analytically such a behavior of the scale factor. Furthermore, once the analytical form of $a(t)$ is found, it is possible to verify the cosmological parameters, namely Hubble and deceleration parameters, which are represented, respectively, by $H(t)=\dot{a}(t)/a(t)$ and $q(t)=-\ddot{a}(t)/\dot{a}(t)H(t)$. 
	
	Under the mathematical perspective, any function that contains approximately all the respective values of $a(t)$ and, at the same time, has the smallest standard deviation, may be considered a suitable analytical expression for the scale factor. The exponential behavior observed in Fig.\ref{NumPlot_a} indicates that a fitting function like $f(t)= A_0 e^{A_1 t}$ should work for the scale factor. However, we shall demonstrate that this is only one of the cases. 
	
	It is important to highlight that for the cases below we use $A_i$, with $i=0,1,2,3...$, as arbitrary constants related to each fitting function. Moreover, the results of each fit come with an individual standard error and its relative deviation from the numerical solution, represented by $\chi$. Therefore, the lower is the value of $\chi$, the higher is the precision of the fitting function in the plot of the numerical solution.	

\subsubsection{An exponential fitting function}

	In the present case we consider a fitting function with a simple exponential form as described by

\begin{equation}\label{scalefit_exp}
a(t) = A_0 e^{A_1 t} +A_2.
\end{equation}	
	
The fitting parameters for  such a function can be seen in Table \ref{t1} below and resulted in
	
\begin{equation}
\chi = 6.566 \times 10^{-10}. \nonumber
\end{equation}
	
\begin{table}[h!]
\center
\begin{tabular}{|c|c|c|}
\hline $A_i$ & $ \bf \beta =-9.4$ & $ \bf \beta =9.4$ \\
\hline $A_0 \times 10^{-3}$ & $ 1.151 \pm 3.894 \times 10^{-4}$ & $ 1.126 \pm 9.684 \times 10^{-5} $ \\
\hline $A_1$ & $ 0.621 \pm 3.484 \times 10^{-5} $ & $ 0.783 \pm 8.847 \times 10^{-6}  $ \\
\hline $A_2 \times 10^{-5}$ & $ -2.991 \pm 5.864 \times 10^{-1} $ & $ -2.747 \pm 4.995 \times 10^{-1} $ \\
\hline
\end{tabular}
\caption{$A_i$ and respective standard errors for the exponential fitting function.}
\label{t1}
\end{table}	



\subsubsection{A polynomial fitting function}

	Since we have observed very small values for the arbitrary constant $A_1$ in Eq. (\ref{scalefit_exp}), we can infer that a polynomial series such as

\begin{equation}\label{scalefit_pol}
a(t) = \sum_{i=0}^9 A_i t^i 
\end{equation}
should satisfy the numerical solution. 

The referred values for the constants involved can be seen in Table \ref{t2} and resulted in

\begin{equation}
\chi =3.896 \times 10^{-10}. \nonumber
\end{equation}

\begin{table}[h!]
\center
\begin{tabular}{|c|c|c|}
\hline $A_i$ & $ \bf \beta =-9.4$ & $\bf \beta =9.4$ \\
\hline $A_0 \times 10^{-4}$ & $ 9.989 \pm 9.206 \times 10^{-3} $ & $ 9.666 \pm 0.259 $ \\
\hline $A_1 \times 10^{-3}$ & $ 1.019 \pm 5.854 \times 10^{-3} $ & $ 1.685 \pm 0.165 $ \\
\hline $A_2 \times 10^{-3}$ & $ -9.815 \times 10^{-2} \pm 1.183 \times 10^{-2} $ & $ -1.773 \pm 0.333 $ \\
\hline $A_3 \times 10^{-3}$ & $ 0.267 \pm 1.056 \times 10^{-2} $ & $ 2.531 \pm 0.297 $ \\
\hline $A_4 \times 10^{-3}$ & $ -9.287 \times 10^{-2} \pm 5.011 \times 10^{-3} $ & $ -1.435 \pm 0.141 $ \\
\hline $A_5 \times 10^{-4}$ & $ 0.302 \pm 1.379 \times 10^{-2} $ & $ 4.975 \pm 0.388 $ \\
\hline $A_6 \times 10^{-6}$ & $ -5.399 \pm 0.227 $ & $ -1.003 \times 10^{2} \pm 6.405 $ \\
\hline $A_7 \times 10^{-7}$ & $ 6.447 \pm 0.221 $ & $ 1.222 \times 10^{2} \pm 6.234 $ \\
\hline $A_8 \times 10^{-8}$ & $ -4.161 \pm 0.117 $ & $ -81.572 \pm 3.300 $ \\
\hline $A_9 \times 10^{-10}$ & $ 13.306 \pm 0.261 $ & $ 2.470 \times 10^{2} \pm 7.322 $ \\
\hline
\end{tabular}
\caption{$A_i$ and respective standard errors for the polynomial fitting function.}
\label{t2}
\end{table}

	The main reason for choosing a ninth degree polynomial function to fit the numerical solution is simply because it stands for the most precision fit. One should expect a high polynomial degree since, as quoted above, an exponential fitting function contains a very low value for $A_1$ such that the expansion in polynomial series is worth. The outcome is clearly an accurate fit, with $\chi$ of the same order of the previous case. However, for this case, we are able to obtain the cosmological time-dependent parameters 
	
\begin{equation}\label{Hpol}
H(t)=\frac{  \sum \limits_{i=0}^9 iA_it^{(i-1)}}{\sum \limits_{i=0}^9 A_i t^i},
\end{equation}

\begin{equation}\label{qpol}
q(t)=-\frac{\sum \limits_{i=0}^9 i(i-1)A_i t^{i-2} \sum \limits_{i=0}^9 A_i t^i}{ \left(\sum \limits_{i=0}^9 iA_i t^{i-1} \right)^2}.
\end{equation}
		
	We can observe the graphical behavior of (\ref{Hpol}) and (\ref{qpol}) in the following Figures \ref{Hpoly_fig}-\ref{qpoly_fig}.

\begin{figure}[h!]
\vspace{0.3cm}
\centering
\includegraphics[height=5cm,angle=00]{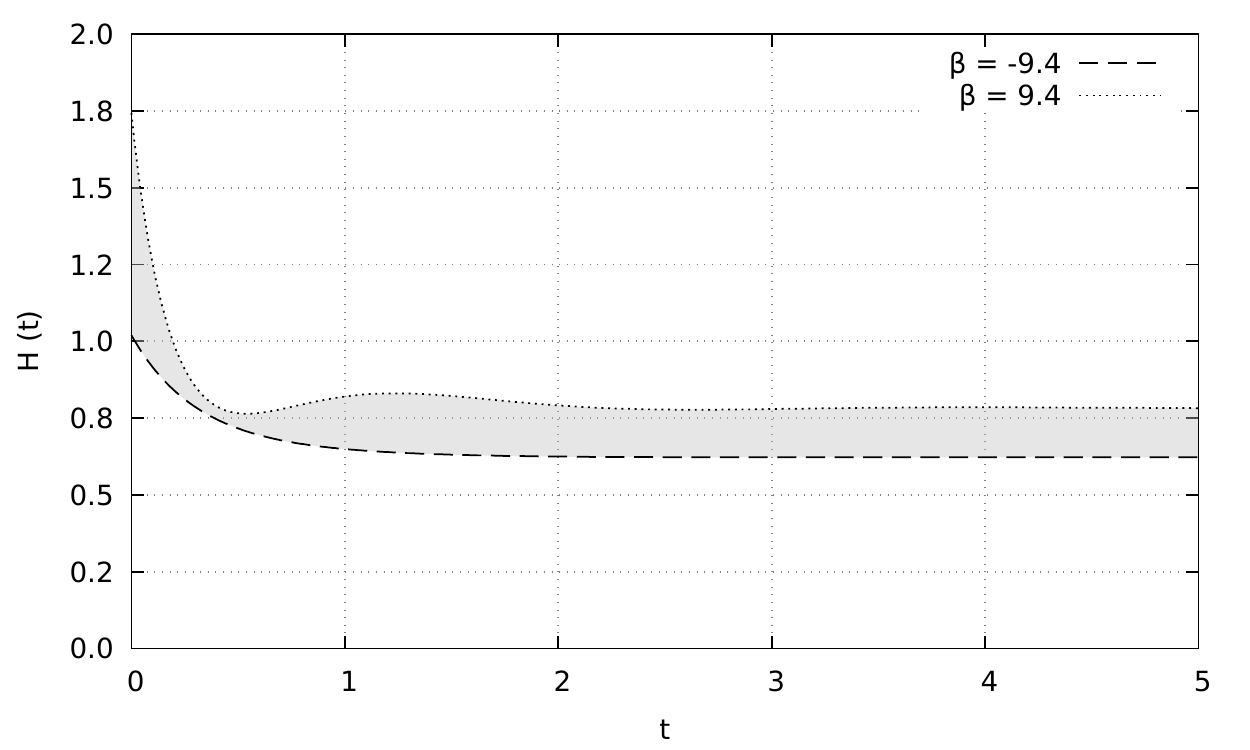}
\caption{Time evolution of the Hubble parameter for the polynomial fit.}
\label{Hpoly_fig}
\end{figure}

\begin{figure}[h!]
\vspace{0.3cm}
\centering
\includegraphics[height=5cm,angle=00]{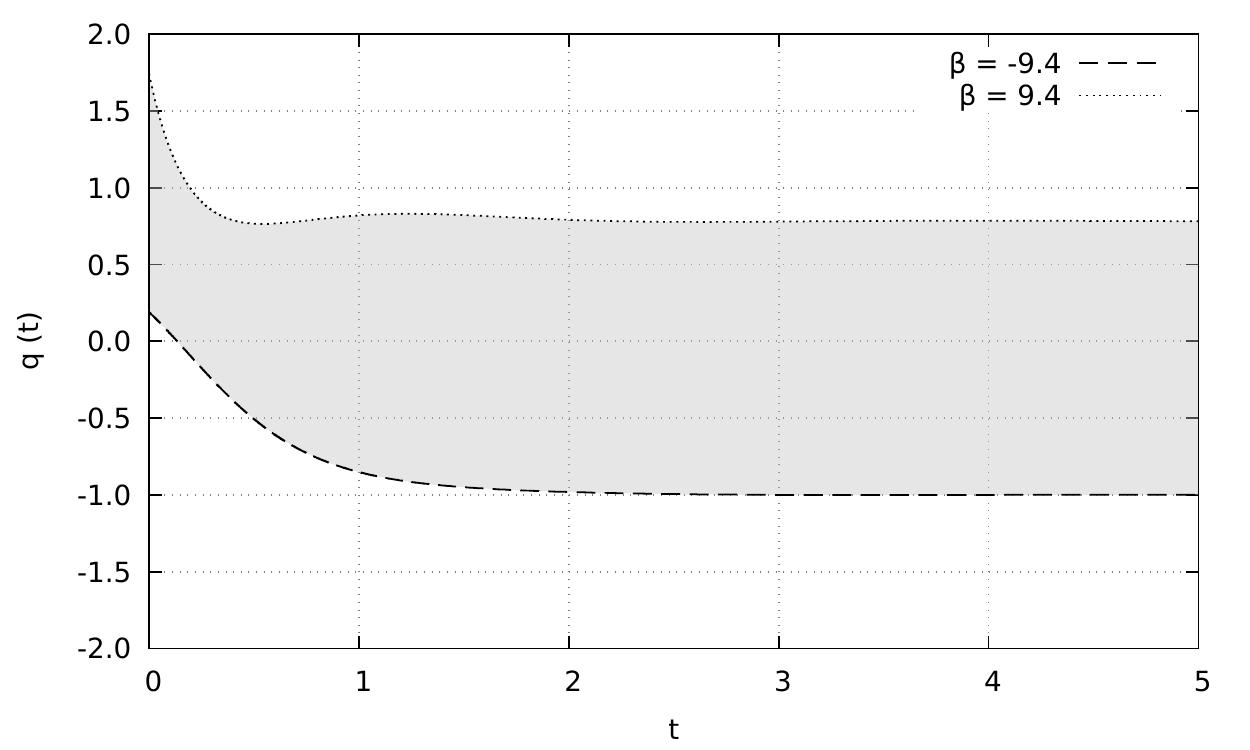}
\caption{Time evolution of the deceleration parameter for the polynomial fit.}
\label{qpoly_fig}
\end{figure}

From Fig.\ref{Hpoly_fig}, the expected behavior $H(t)\sim t^{-1}$ is clearly observed. In Fig.\ref{qpoly_fig} one can notice the transition from a decelerated ($q>0$) to an accelerated ($q<0$) phase of the Universe expansion for $\beta=-9.4$.

\subsubsection{A trigonometric fitting function}

Here we will present another fitting candidate, namely a trigonometric function, expressed by 

\begin{equation}\label{scalefit_tri}
a(t)=A_0 \sinh(A_1 t)+A_2.  
\end{equation} 
As it is known, hyperbolic functions can be described as combinations of exponentials. Therefore, such a fitting function should be also intuitive suchlike (\ref{scalefit_exp}). The values found for the free parameters can be found in Table \ref{t3} and yield

\begin{equation}
\chi = 3.943 \times 10^{-8}. \nonumber
\end{equation}

\begin{table}[h!]
\center
\begin{tabular}{|c|c|c|}
\hline $A_i$ & $ \bf \beta =-9.4$ & $ \bf \beta =9.4$ \\
\hline $A_0 \times 10^{-3}$ & $ 2.269 \pm 5.607 \times 10^{-3} $ & $ 2.245 \pm 1.561 \times 10^{-3} $ \\
\hline $A_1$ & $ 0.624 \pm 2.547 \times 10^{-4} $ & $ 0.783 \pm 7.151 \times 10^{-5} $ \\
\hline $A_2 \times 10^{-4}$ & $ 4.043 \pm 0.452 $ & $ 2.631 \pm 0.403 $ \\
\hline
\end{tabular}
\caption{$A_i$ and the respective standard errors for the trigonometric fitting function.}
\label{t3}
\end{table}	

From such a trigonometric function, it is possible to see a precise fit, with low error values for the constants $A_i$ and, most importantly, a small deviation from the numerical solution as it can be seen from the $\chi$ value. 

This model yields the following cosmological parameters:

\begin{equation}\label{Htri}
H(t) = \frac{A_0 A_1 \cosh(A_1 t)}{A_0 \sinh(A_1 t)+A_2},
\end{equation}
\begin{equation}\label{qtri}
q(t) = -\frac{\sech(A_1 t) \tanh(A_1 t) \left[A_0 \sinh(A_1 t)+A_2 \right]}{A_0}.
\end{equation}

\begin{figure}[h!]
\vspace{0.3cm}
\centering
\includegraphics[height=5cm,angle=00]{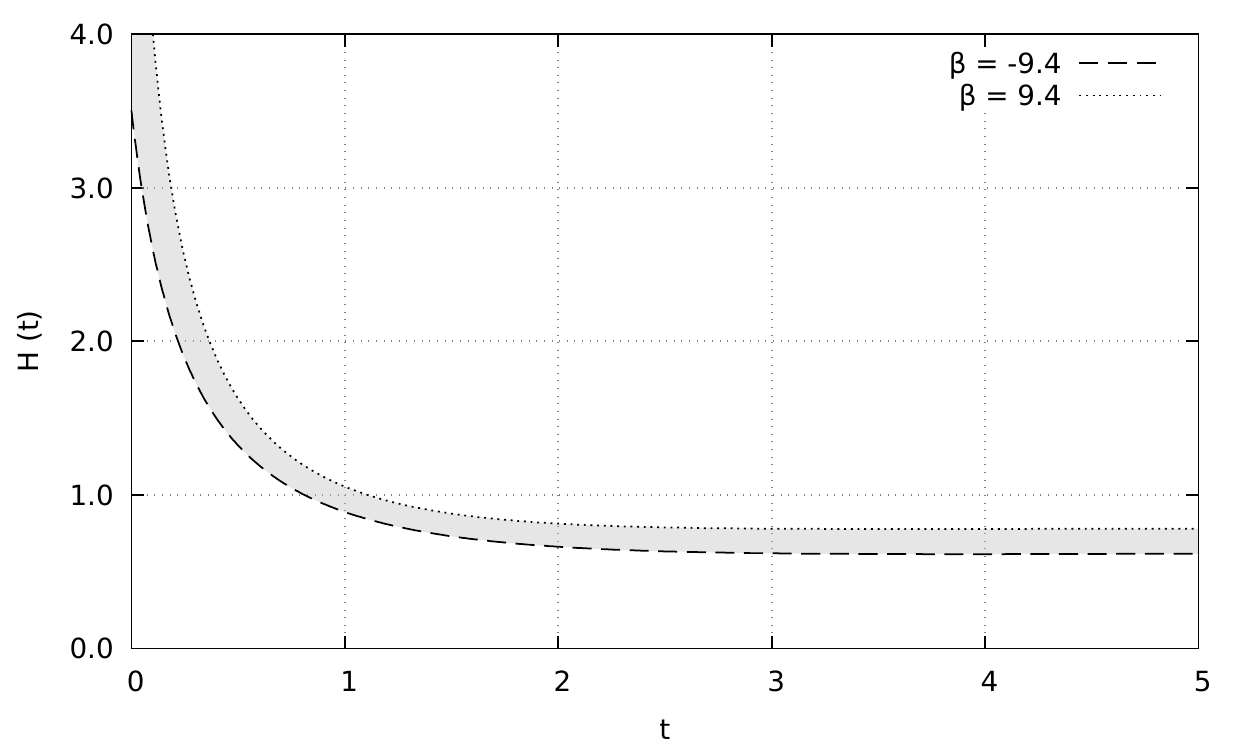}
\caption{Time evolution of the Hubble parameter from the trigonometric fit.}
\label{Htrig_fig}
\end{figure}

\begin{figure}[h!]
\vspace{0.3cm}
\centering
\includegraphics[height=5cm,angle=00]{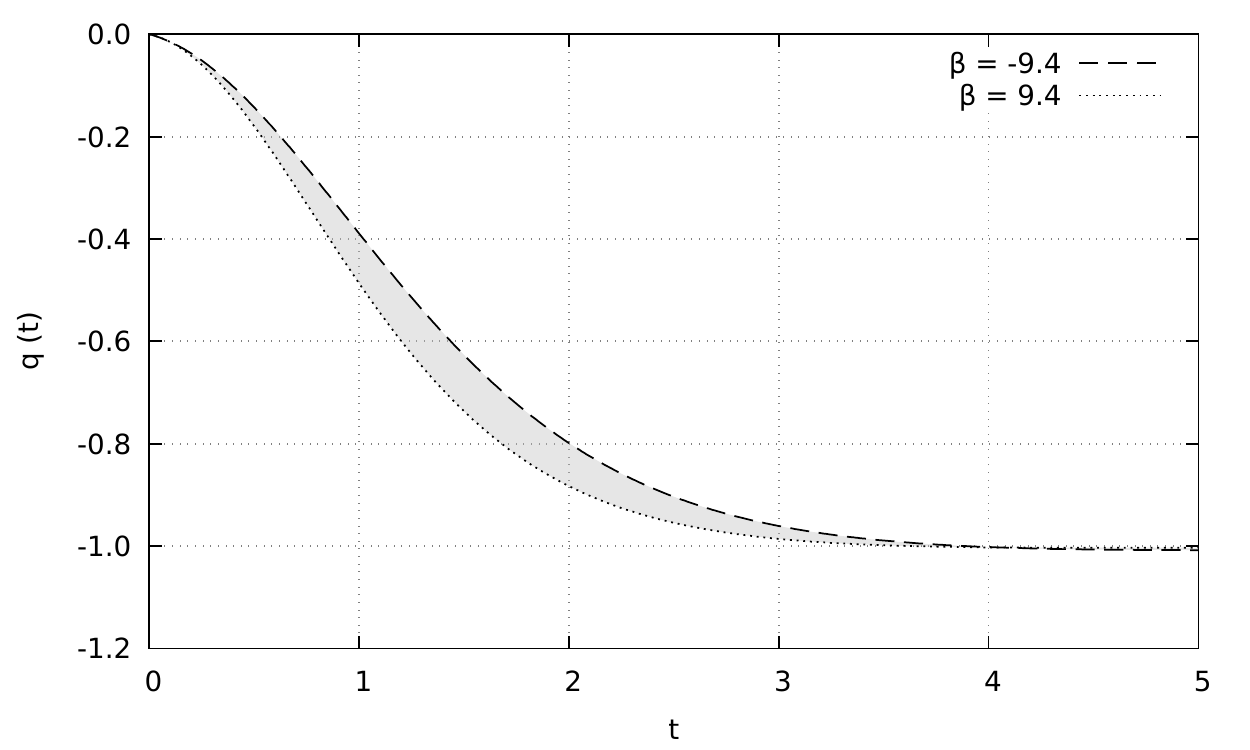}
\caption{Time evolution of the deceleration parameter from the trigonometric fit.}
\label{qtrig_fig}
\end{figure}

The evolution of such quantities in time can be seen in Figs.\ref{Htrig_fig}-\ref{qtrig_fig}.

\subsubsection{A power law fitting function}
	
	In this case we fit our numerical solution for the scale factor with a power law function such as 

\begin{equation}\label{scalefit_pow}
a(t)=A_0 t^{A_1} +A_2.
\end{equation}
Such a function is normally seen in the text books of cosmology \cite{ryden/2003}, to describe the different dynamical stages of the Universe evolution, each of them for a different value of $A_1$.

The values obtained for the referred constants can be seen in Table	\ref{t4} below and have yielded to

\begin{equation}
\chi = 2.457\times10^{-5}. \nonumber
\end{equation}

\begin{table}[h!]
\center
\begin{tabular}{|c|c|c|}
\hline $A_i$ & $ \bf \beta =-9.4$ & $ \bf \beta =9.4$ \\
\hline $A_0 \times 10^{-7}$ & $ 1.559 \times 10 \pm 2.317 $ & $ 2.210 \pm 0.345 $ \\
\hline $A_1 $ & $ 5.554 \pm 6.540 \times 10^{-2} $ & $ 7.095 \pm 6.831 \times 10^{-2} $ \\
\hline $A_2 \times 10^{-3}$ & $ 7.861 \pm 1.083 $ & $ 21.431 \pm 3.899 $ \\
\hline
\end{tabular}
\caption{$A_i$ and respective standard error of the power law fitting function.}
\label{t4}
\end{table}	

For this case, the cosmological parameters are analytically described by 
	
\begin{equation}\label{Hpow}
H(t)= \frac{A_0 A_1 t^{(A_1-1)}}{A_0 t^{A_1}+A_2},
\end{equation}
\begin{equation}\label{qpow}
q(t)=- \frac{(A_1 -1)(A_0 t^{A_1}+A_2)t^{-A_1}}{A_0 A_1}.
\end{equation}

The graphical behavior of these quantities can be seen in Figs.\ref{Hgeo_fig}-\ref{qgeo_fig} below.

\begin{figure}[h!]
\vspace{0.3cm}
\centering
\includegraphics[height=5cm,angle=00]{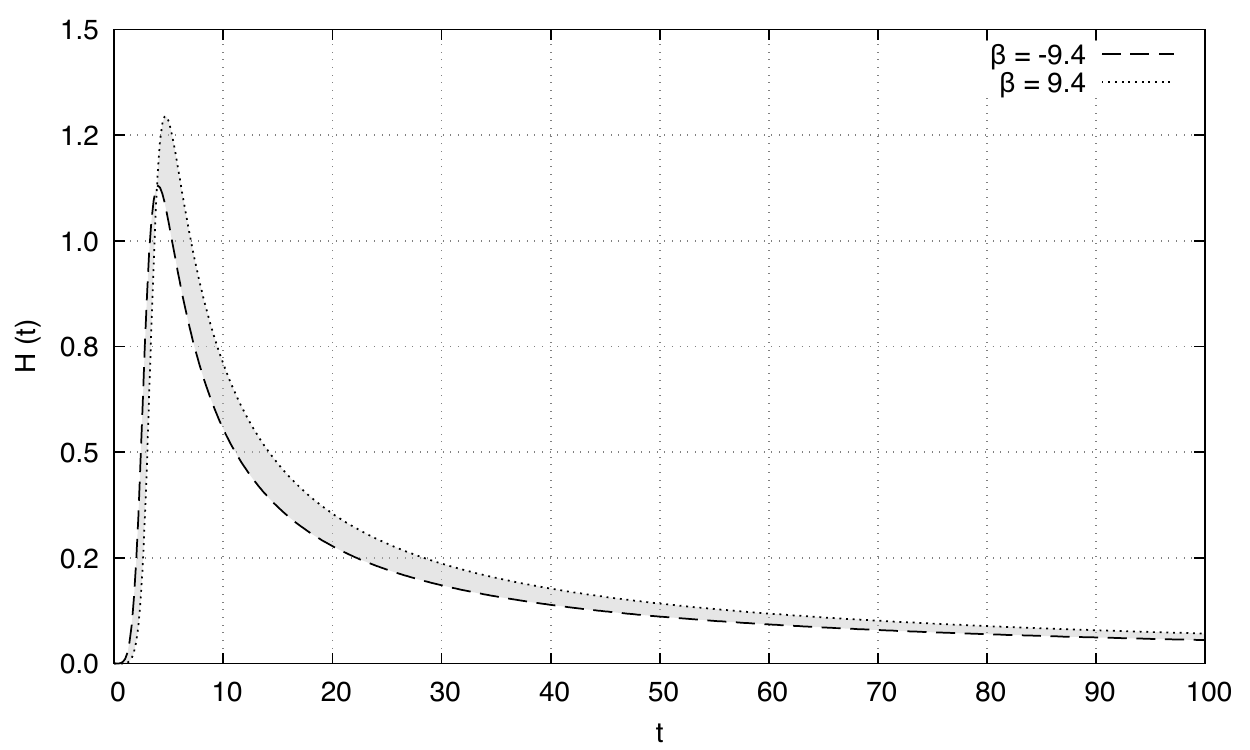}
\caption{Time evolution of the Hubble parameter from the power law fit.}
\label{Hgeo_fig}
\end{figure}

\begin{figure}[h!]
\vspace{0.3cm}
\centering
\includegraphics[height=5cm,angle=00]{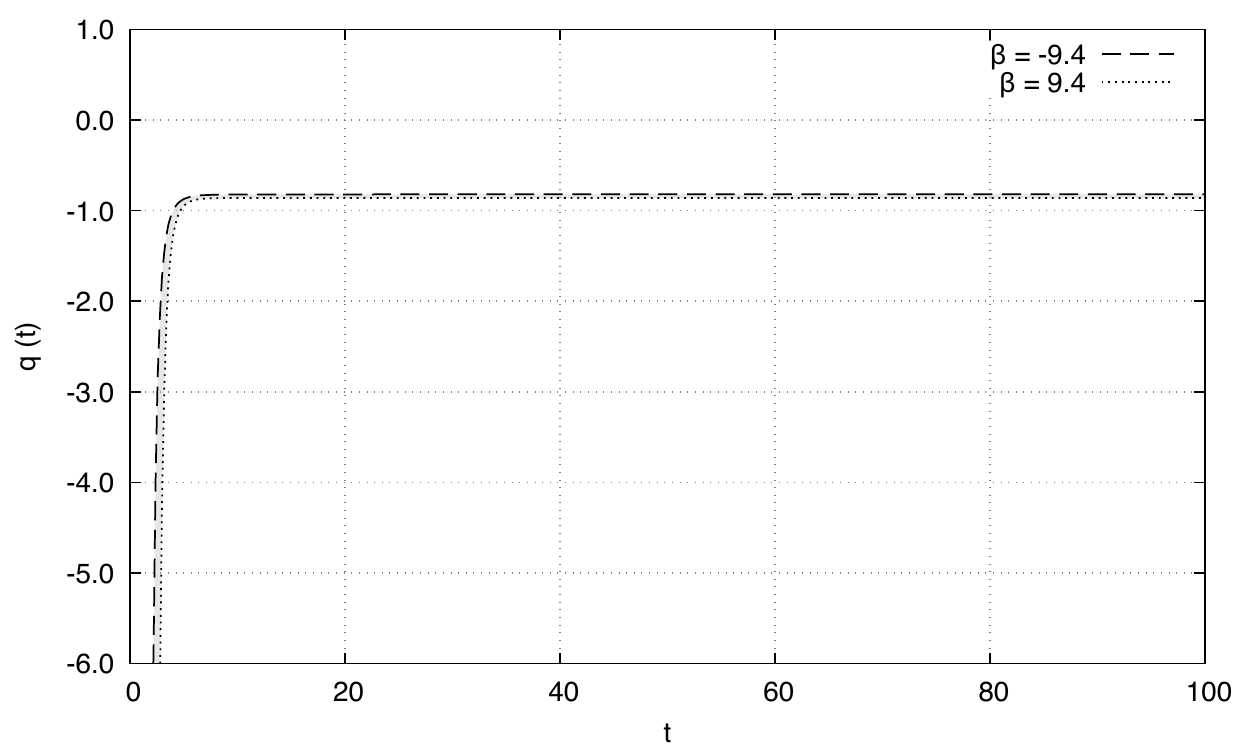}
\caption{Time evolution of the deceleration parameter from the power law fit.}
\label{qgeo_fig}
\end{figure}

One may observe unusual behaviors in both figures above. In Fig.\ref{Hgeo_fig} for $t \geq 4 $, the well expected decreasing behavior of $H(t)$ is obtained, as well as a constant value for high values of time, which should be expected in an accelerated expanding Universe, as we are going to revisit below. However, for $t<4$ the Universe decreases its length scales, which could indicate a bounce cosmological model \cite{brandenberger/2017,battefeld/2015}. Anyhow, we should discard this model based on Fig.\ref{qgeo_fig}, which presents an always-accelerated expanding Universe, which strongly departs from the structure formation scenario \cite{ryden/2003}.

\section{Exponential dependence and scalar field cosmology}
\label{sec_3}

In this section we use the coupling between a real scalar field Lagrangian with the previous $f(R,T)$ function in order to find analytical cosmological scenarios.  If we consider a standard scalar field Lagrangian given by
\begin{equation}
{\cal L}_{m}={\cal L}=\frac{1}{2}\,\partial_{\mu}\phi\partial^{\mu}\phi-V(\phi)\,,
\end{equation}
where $\phi=\phi(t)$ also known as inflaton field. By working with such a Lagrangian, we have the following components of the energy-momentum tensor
\begin{equation}
\rho=\frac{\dot{\phi}^2}{2}+V(\phi)\,,
\end{equation}
and
\begin{equation}
p=\frac{\dot{\phi}^2}{2}-V(\phi)\,.
\end{equation}
So, taking these previous ingredients into the Friedmann Eqs. $(\ref{00})$, and $(\ref{11})$ we yield to
\begin{eqnarray} \label{friedmann01}
H^{\,2}&=&\frac{8\,\pi}{3}\,\left(\frac{\dot{\phi}^{\,2}}{2}+V\right) \\ \nonumber
&&
+\alpha\,\left[\frac{1}{2}-\beta\,\left(\dot{\phi}^{\,2}-4\,V\right)\right]\,e^{\,-\beta\,(\dot{\phi}^{\,2}-4\,V)}\,,
\end{eqnarray}
and
\begin{equation} \label{friedmann02}
\dot{H}=-4\,\pi\,\dot{\phi}^{\,2}+\alpha\,\beta\,\left(\dot{\phi}^{\,2}-4\,V\right)\,e^{\,-\beta\,(\dot{\phi}^{\,2}-4\,V)}\,.
\end{equation}
One can see that the standard cosmological scenario in the presence of a background scalar field is recovered if $\alpha=0$. Moreover, the time-dependent scalar field $\phi(t)$ must obey the equation of motion
\begin{equation}
\left(1-2\,f^{\,\prime}\right)\,\left(\ddot{\phi}+3\,H\,\dot{\phi}\right)-2\,\dot{f}^{\,\prime}\,\dot{\phi}+\left(1-4\,f^{\,\prime}\right)\,V_{\,\phi}=0\,,
\end{equation}
which can be rewritten as
\begin{eqnarray}
&&
\ddot{\phi}+3\,H\,\dot{\phi}+V_{\,\phi}-2\,\alpha\,\beta\,e ^{\,-\beta\,(\dot{\phi}^{\,2} -4\,V)}\\ \nonumber
&&
\times\,\left[\left(1+\beta\,\dot{\phi}\right)\,\ddot{\phi}+3\,H\,\dot{\phi}-2\,\left(1-\beta\,\dot{\phi}\right)\,V_{\,\phi}\right]=0\,.
\end{eqnarray}

\subsection{Slow-roll approximation}

One way to find analytical behaviors for the cosmological parameters consists in work with the slow-roll regime, where we consider the following approximations
\begin{equation}\label{conditions}
\ddot{\phi}=0\,; \qquad \dot{\phi}^{\,2} << V\,; \qquad \beta \approx 0\,,
\end{equation}
therefore, Eqs. $(\ref{friedmann01})$, and $(\ref{friedmann02})$ are simply given by
\begin{equation} \label{fd01}
H^{\,2}=\left(\frac{8\,\pi}{3}+6\,\alpha\,\beta\right)\,V+\frac{\alpha}{2}\,,
\end{equation}
and
\begin{equation} \label{fd02}
\dot{H}=-4\,\pi\,\dot{\phi}^{\,2}\,.
\end{equation}
Moreover, the equation of motion for the scalar field is reduced to
\begin{equation} \label{eqm01}
3\,\left(1-2\,\alpha\,\beta\right)\,H\,\dot{\phi}=-\left(1+4\,\alpha\,\beta\right)\,V_{\,\phi}\,.
\end{equation}

By combining the previous results we yield to the following constraint 
\begin{equation}
\frac{1+4\,\alpha\,\beta}{3\,\left(1-2\,\alpha\,\beta\right)}=\frac{1}{3}+\frac{3\,\alpha\,\beta}{4\,\pi}\,,
\end{equation}
which imposes the solutions
\begin{equation}
\alpha=0\,; \qquad \alpha=\frac{3-8 \pi }{6 \,\beta }\,.
\end{equation}
We can see that the solution $\alpha=0$ recovers the standard slow-roll approximation for general relativity, while the other one allows us to account for new effects, which are related with the trace of the energy-momentum tensor. 
Now, let us find some physical cosmological parameters by taking a Klein-Gordon potential, which means
\begin{equation}
V=\frac{m^{\,2}}{2}\,\phi^{\,2}+V_0\,,
\end{equation}
where $m$, and $V_0$ are real constants. So, $(\ref{fd01})$ results in
\begin{equation} \label{fd03}
H^{\,2}=\frac{8\,\pi-3}{12\,\beta}+\frac{\left(9-16\,\pi\right)}{6}\,\left(m^{\,2}\,\phi^{\,2}+2\,V_0\right)\,,
\end{equation}
where in the last equation we changed $\beta\rightarrow -\beta$ in order to derive real parameters.  Then, taking the Hubble parameter into the equation of motion $(\ref{eqm01})$, and considering an expansion around $\beta\approx 0$, we obtain
\begin{equation}
\dot{\phi}\approx c_1\,\phi^{\,2}\,; \qquad c_1=-\frac{(16 \,\pi -9) \sqrt{\beta }}{4\, \pi\,  \sqrt{24 \,\pi -9}}\,m^{\,2}\,,
\end{equation}
whose solution is
\begin{equation}
\phi(t) =-\frac{1}{c_1 t+c_2}\,.
\end{equation}

\begin{figure}[h!]
\vspace{0.3cm}
\centering
\includegraphics[height=5cm,angle=00]{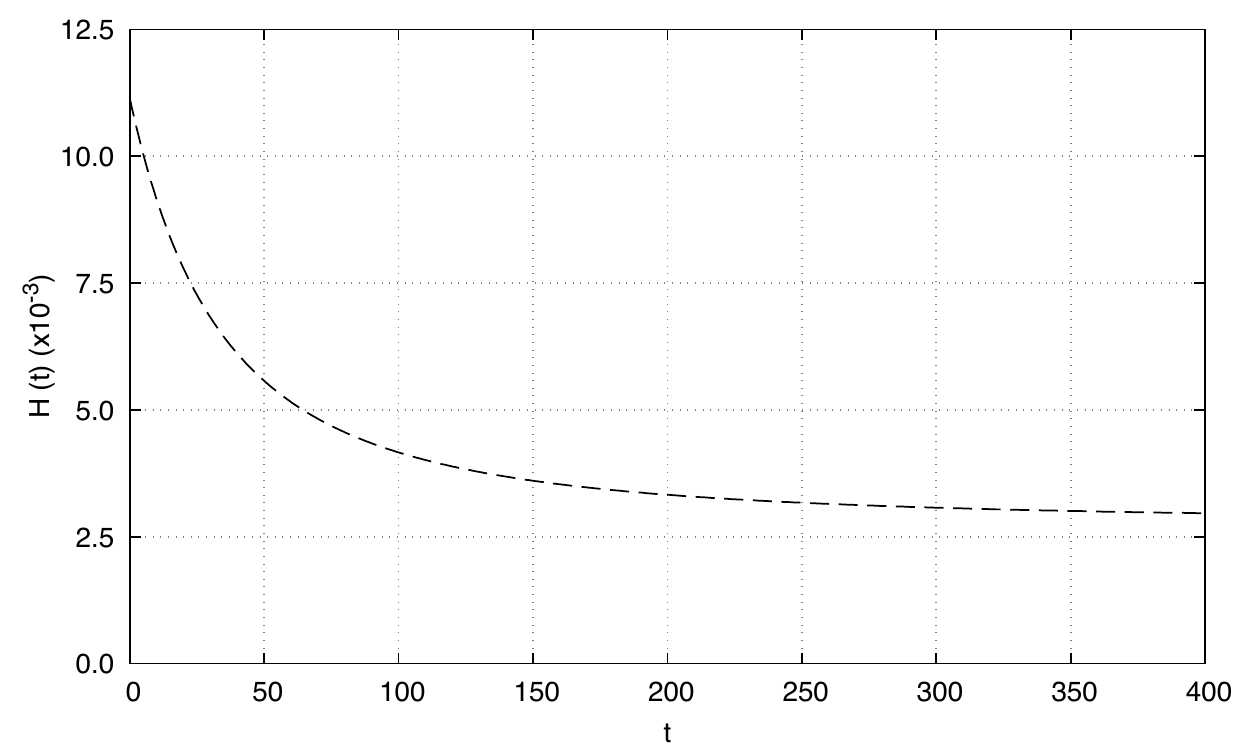}
\caption{Time evolution of the Hubble parameter for the slow-roll approximation. In this graphic we considered $c_2=-11$, $\beta=0.01$, $V_0=26.823$, and $m=2$.}
\label{H_slowroll}
\end{figure}

\begin{figure}[h!]
\vspace{0.3cm}
\centering
\includegraphics[height=5cm,angle=00]{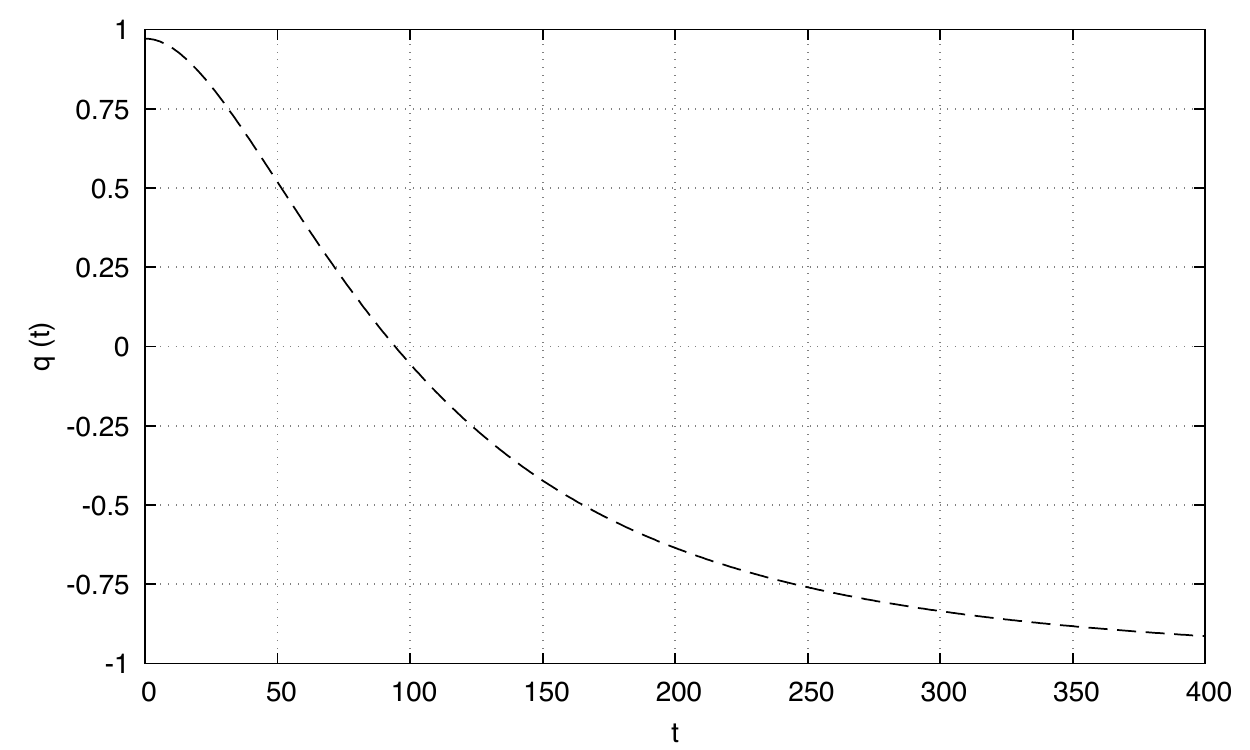}
\caption{Time evolution of the deceleration parameter for the slow-roll approximation. In this graphic we considered $c_2=-11$, $\beta=0.01$, $V_0=26.823$, and $m=2$.}
\label{q_slowroll}
\end{figure}

Let us go back to $(\ref{fd03})$, and rewrite it as 
\begin{equation}
H(t)\approx \frac{(16 \pi -9) \sqrt{\beta } \left(m^2 \phi ^2+2\,V_0\right)}{2 \sqrt{24 \pi -9}}-\frac{\sqrt{\frac{8 \pi }{3}-1}}{2 \sqrt{\beta }}\,,
\end{equation}
where we consider an expansion around $\beta \approx 0$. Therefore, taking the field $\phi(t)$ into the last expression yields to
\begin{equation}
H(t)=\frac{(16 \pi -9) \beta\,\left(2\,V_0+\frac{ m^2}{\left(c_2+\frac{(9-16 \pi ) \sqrt{\beta } m^2 t}{4 \pi  \sqrt{24 \pi -9}}\right)^2}\right)+3-8 \pi}{2 \sqrt{24 \pi -9} \sqrt{\beta }}\,,
\end{equation}
representing an analytical form for the Hubble parameter. The features of $H(t)$ can be appreciated in Fig. \ref{H_slowroll}, which shows that the Hubble parameter is decaying as time passes by. Moreover, after a long period, this parameter becomes approximately a positive constant, such a feature corroborates with the actual phase of accelerated expansion.  As a matter of completeness, we also depicted the graphic for the deceleration parameter, which is shown in Fig. \ref{q_slowroll}. We also would like to point out that we did not include the analytical expression for $q$ because of its size.

As one can see, the previous parameter starts in $q\approx 1$, which is expected at the radiation era, after that, the parameter gradually evolves to negative values. As time passes by, $q$ goes to $0.5$ corresponding to the matter era, and finally decays to $-1$ which is consistent with the dark energy regime. 

\section{Discussion and Conclusion}

In this paper, we have proposed an alternative model for $f(R,T)$ theory of gravity, based on an exponential function of the trace. Although the $\Lambda$CDM as a cosmological model shows agreement with current observations, it also presents theoretical shortcomings \cite{weinberg/1989,wang-yang/2006}, where the need for an exotic matter is claimed and shallowly understood \cite{adamek/2016}.

The $f(R,T)$ theory stands with wide analysis and application, suchlike compact massive objects \cite{bhatti/2017,mcl/2017}, anisotropic and homogeneous metrics \cite{reddy/2012,sharif-zubair/2014}, perspective through the inflationary era \cite{msrc/2019}, among other investigations. Therefore, the modified gravity express a viable path on cosmology, once it is in agreement with observable parameters, besides it can be applied to several interesting scenarios.

Introducing the modified theory of $f(R,T)$ \cite{harko/2011} and its viability in section I, where $R$ is the Ricci scalar and $T$, the energy-momentum trace, we proposed an unprecedented algebraic function to current literature. It is relevant to inform that an exponential dependence was investigated only for $f(R)$ theories, as we can see in \cite{cognola/2008}. Taking the usual FLRW metric, the mathematical implication of using this expression in the Einstein-Hilbert action of gravity is carefully analyzed in section II. The Friedmann equations (\ref{00}), and (\ref{11}) are presented with a highly non-linear behavior for such a case. Herewith we use the intuitive EoS of matter regime ($\omega = 0$), where a null pressure is considered.

The consideration of a null pressure EoS leads to a numerical method on finding the solution for the scale factor, integrating the Friedmann equations. In Fig. \ref{NumPlot_a} we plotted the numerical integrations of the scale factor, taking appropriate conditions on both $\alpha$ and $\beta$ parameters. This graphic also presented a wide range of valid solutions for $a(t)$. Moreover, Fig. \ref{NumPlot_a} unveils that $a(t)$ generates a nice cosmic growth, which is extremely suitable from the cosmological framework. To obtain the analytical expression from the numerical approach, we have declared several fits, either by mathematical or physical convenience on cosmology. The precision of each fitting function is analyzed by its respective $\chi$ value.

An exponential fit is highly intuitive, expressing the most precise fitting function, and whose parameters are presented in Table \ref{t1} for $\chi \sim 10^{-10}$. Another possible path for fitting $a(t)$ is through a polynomial function, as shown in Eq. \eqref{scalefit_pol}, where the sum is stopped at $i=9$ for a matter of precision. We have derived the Hubble and deceleration parameters for this case, which are presented in Figs. \eqref{Hpoly_fig} and \eqref{qpoly_fig}, respectively. These graphics show not only agreement with both theoretical and observational data \cite{perlmutter/1999,beringer/2012}, but a wide range of validity in which $\beta\in[-9.4,+9.4]$. Following our numerical approach, we made also a trigonometric fit which yields us to nice and precise values for the fit candidates, as one can see in Table \ref{t3} for $\chi \sim 10^{-8}$. In this scenario, we find a well-behaved transition from null to accelerated cosmic phase, whose features are presented in Fig. \ref{qtrig_fig}. Such a transition corroborates with predictions of a de-Sitter description for the Universe \cite{zhang/2011,cognola/2009,setare-moh/2012}. Last but not least we have used a power-law fitting function, as shown in Equation \eqref{scalefit_pow}. This function is commonly used in standard cosmology as one can see in \cite{ryden/2003}. For this fit, we find the least precise function, whose parameters are shown in Table \ref{t4} for $\chi \sim 10^{-5}$. Furthermore, although the deceleration parameter recovers the de-Sitter Universe ($q\rightarrow -1$), Fig. \ref{qgeo_fig} indicates that the Universe has been always under accelerated expansion. Meanwhile from Fig. \ref{Hgeo_fig} we observe a noteworthy bounce, leading to cycling cosmological model \cite{cham-muk/2017,nojiri/2017}.

As another test for the viability of this new proposal for $f(R,T)$ gravity, we investigated the cosmological parameters derived from its coupling with a real scalar field (or inflaton field). There, we were able to find the Friedmann equations as well as the equation of motion for the scalar field. Once such differential equations were highly non-linear we were impelled to use the slow-roll approximation to derive analytical cosmological parameters. The cosmological scenario produced can be seen in Figs. \ref{H_slowroll} and \ref{q_slowroll}, and corroborates with the actual phase of accelerated expansion our Universe passes through. The behaviors of the analytical parameters are in agreement with our numerical tests for polynomial and trigonometric fits, as one can see in Figs. \ref{Hpoly_fig} - \ref{qtrig_fig}. Moreover, in Fig. \ref{q_slowroll} we presented a nice continuous transition between matter ($q \sim 0.5$) and dark energy ($q \sim -1$) regimes, reproducing the outcomes from numerical calculations with polynomial fit. 

\section*{Acknowledgements}

G. Ribeiro would like to thank the Brazilian National Research Council (CNPq) for providing a scholarship program. R. Sfair would like to thank S\~ao Paulo Research Foundation (FAPESP), grants 2016/24561-0 for financial support. P.H.R.S. Moraes would like to thank S\~ao Paulo Research Foundation (FAPESP), grants
2018/20689-7, and 2015/08476-0, for financial support. J. R. L. Santos would like to thank CNPq for financial support, grant 420479/2018-0, and CAPES.

\end{document}